\documentclass[reprint,aps,pra,amsmath,amssymb]{revtex4-1}

\usepackage{setspace}
\usepackage{graphicx}
\usepackage{amsmath}
\usepackage{lipsum}
\usepackage{cases}
\usepackage{longtable}
\usepackage{afterpage}
\usepackage{rotating}
\usepackage{threeparttable}
\usepackage{booktabs}
\usepackage{multirow}
\usepackage{subfigure}
\usepackage{array}
\usepackage{color}
\usepackage{float}
\usepackage[section]{placeins}
\usepackage{longtable}
\usepackage{nameref}
\usepackage{bm}
\usepackage{lipsum}
\usepackage{appendix}
\usepackage{soul}

\begin{document}

\title{Revisiting the hyperfine interval for the $2s2p$ $^3\!P_{J}$ state in $^9$Be}
\author{Yu-Shan Zhang,$^{1,2}$ Wei Dang,$^{1,*}$ Kai Wang,$^{3,*}$ and Yong-Bo Tang$^{2,*}$}

\affiliation {$^1$Hebei Key Lab of Optic-electronic Information and Materials, The College of Physics Science and Technology, Hebei University, Baoding 071002, China}
\affiliation {$^2$Physics Teaching and Experiment Center, Shenzhen Technology University, Shenzhen, 518118, China}
\affiliation {$^3$Department of Physics, Anhui Province Key Laboratory for Control and Applications of Optoelectronic Information Materials, Key Laboratory of Functional
Molecular Solids, Ministry of Education, Anhui Normal University, Wuhu, Anhui, 241000, China}
\email{tangyongbo@sztu.edu.cn}
\email{$\rm wang_kai10@fudan.edu.cn$}
\email{dangwei@hbu.edu.cn}
\date{\today}

\begin{abstract}
Using relativistic multiconfiguration Dirac-Hartree-Fock method, we calculate the hyperfine-structure properties of the $2s2p$ $^3\!P_{J}$ state in $^9$Be. The hyperfine-structure properties encompass first-order hyperfine-structure parameters, as well as second-order and third-order corrections arising from the hyperfine mixing of different $2s2p$ $^3\!P_{J}$ levels. Based on our theoretical results, we reanalyze the previously reported measurement of the hyperfine interval for the $2s2p$ $^3\!P$ state in $^9$Be [A. G. Blachman and A. Lurio, Phys. Rev. 153, 164(1967)], yielding updated hyperfine-structure constants. Our results show that the hyperfine-structure constant $B$ of $2s2p$ $^3\!P_{1}$ is notably sensitive to second-order correction. Conversely, accurately determining the hyperfine-structure constant $B$ of $2s2p$ $^3\!P_{2}$ necessitates consideration of the hyperfine-structure constant $C$ in the first-order hyperfine interaction equation. The updated hyperfine-structure constant $B$ of the $2s2p$ $^3\!P_{2}$ state is found to be $1.4542(67)$~MHz, which is approximately $1.7\%$ larger than the previous value of $1.427(9)$~MHz.
 By combining our theoretical results with the updated hyperfine-structure constant for the $2s2p$ $^3\!P_{2}$ state, we extract the electric quadrupole moment $Q$ of $^9$Be nucleus to be $0.05320(50)$~b. This value is consistent with the most recent determination using the few-body precision calculation method. Additional, we also discuss
the reasons for the discrepancy between the $Q$ values obtained through few-body and previous many-body calculations.
\end{abstract}
\pacs{31.15.ac, 31.15.ap, 34.20.Cf}
\maketitle
\section{Introduction}
The electric quadrupole moment $Q$ is a fundamental parameter that quantifies the deformation of the nuclear charge distribution relative to spherical symmetry. Accurate knowledge of this quantity is essential for enhancing our understanding of nucleon-nucleon interactions~\cite{Cocolios_2009_prl,Yordanov_2013_prl,Papuga_2013_prl}. Although many nuclei have well-established benchmark values for their magnetic dipole moments, precise reference values for the electric quadrupole moments of numerous nuclei remain inadequate~\cite{Stone_2005_adnd,Pekka2017}. While theoretically, nuclear multipole moments can be evaluated using nuclear model theory, the accuracy of this method is heavily dependent on the specific nuclear model employed. A more model-independent approach to determining the electromagnetic multipole moments of the nucleus involves integrating the hyperfine structure spectrum with corresponding high-precision atomic or molecular calculations. Presently, this is one of the most accurate methods for determining the electric quadrupole moment $Q$ of both heavy and unstable nuclei~\cite{Singh2012pra,Sahoo2013pra,Bi2018pra,Lu2019pra,Li2021pra,li2021ab,Porsev2021prl,Papoulia2021pra,Skripnikov2021prc,Skripnikov2024prc}.

More than 50 years ago, Blachman and Luris~\cite{blachman1967hyperfine} precisely measured the hyperfine splitting of the $2s2p$ $^3\!P_J$ states in the $^9$Be atom using the atomic-beam magnetic resonance method. After incorporating second-order and third-order corrections due to hyperfine mixing among the $2s2p$ $^3\!P_J$ fine energy levels, they deduced the magnetic dipole and electric quadrupole hyperfine-structure (HFS) constants for $2s2p$ $^3\!P_{1}$ and $2s2p$ $^3\!P_{2}$ states, and additionally determined the electric quadrupole moment $Q$ of $^9$Be nucleus. However, their estimation of $Q$ carried a $6\%$ uncertainty due to the limited precision in calculated hyperfine interaction matrix elements. Subsequently, some research groups recalculate the hyperfine interaction parameters of $^9$Be using high-precision many-body methods, and redefined the electric quadrupole moment $Q$ of $^9$Be nucleus by combining the experimental HFS $B$ for the $2s2p$ $^3\!P_2$ state\cite{ray1973study,sinano1973hfs,beck1984fine,sundholm1991large,jonsson1993large,nemouchi2003theoretical,beloy2008hyperfine}. For example, Ray \textit{ et. al.}\cite{ray1973study} calculated the electric field gradient of the $2s2p$ $^3\!P_{2}$ state using linked-cluster many-body perturbation theory, obtaining $Q = 0.0525(3)$~b. Sundholm and Olsen\cite{sundholm1991large} employed the multiconfiguration Hartree-Fock (MCHF) method to calculate the electric field gradient of the $2s2p$ $^3\!P_{2}$ state, resulting in $Q = 0.05288(38)$~b. J\"{o}nsson and Fischer performed a large-scale  multiconfiguration Hartree-Fock (MCHF) calculation of HFS constants and got $Q = 0.05256$~b.~\cite{jonsson1993large}.
These theoretical studies, conducted within a non-relativistic framework, showed consistency and significantly improved accuracy over Blachman and Luris's result. Notably, $Q = 0.05288(38)$~b is currently the recommended value~\cite{Pekka2017}. In 2006, Beloy \textit{ et. al.} conducted a relativistic calculation of the magnetic dipole, electric quadrupole, and magnetic octupole HFS constants for the $2s2p$ $^3\!P_2$ in $^9$Be using relativistic configuration interaction plus second-order many-body perturbation theory. Although they reported a $Q$ value of $0.053(3)$~b, their primary focus was on assessing the contribution of second-order corrections due to off-diagonal hyperfine interactions to the magnetic octupole HFS constant. When determining the $Q$ value, the aforementioned works directly utilized the experimental HFS $B$ for the $2s2p$ $^3\!P_2$ state as reported by Blachman and Luris.

Recently, Puchalski \textit{ et. al.}\cite{puchalski2021hyperfine} performed highly accurate calculations of the hyperfine structure of the $2s2p$ $^3\!P$ state in $^9$Be with the help of highly optimized explicitly correlated Gaussian functions. Their calculations considered the leading finite nuclear mass, radiative, nuclear structure, and relativistic effects. For convenience, we will hereafter refer to their method as the ''few-body precision calculation method". By comparing their results with the measurement by Blachman and Luris\cite{blachman1967hyperfine}, they extracted the electric quadrupole moment $Q$ to be $0.05350(14)$~b. This value, the most accurate to date, differs from the currently recommended value of $Q = 0.05288(38)$~b.   Accurate knowledge of the electric quadrupole moment $Q$ of $^9$Be is very important for precision spectroscopies and accuracy calculations on $^9$Be atom and ions~\cite{Qi2023pra,Qi2024pra,Fairbank2024pra,Dickopf2024nature}.
Previous determinations of the $Q$ value were based on the calculations by many-body methods. Be is a four-electron atomic system, many-body methods such as multiconfiguration Hartree-Fock, multiconfiguration Dirac-Fock, configuration interaction, and coupled cluster methods, are theoretically capable of accurately calculating hyperfine interaction properties of Be. Therefore, further investigation into the reasons for the discrepancy between the $Q$ values obtained through few-body precision calculation method and many-body calculations is essential.

In the present work, we perform a relativistic calculation of the hyperfine interaction properties of $2s2p$ $^3\!P_{J}$ states in the $^9$Be atom. We first apply the multiconfiguration Dirac-Hartree-Fock (MCDF) and relativistic configuration interaction (RCI) methods to obtain the wave-functions of $2s2p$ $^3\!P_{J}$ states. Subsequently, we use these wave-functions to calculate the magnetic dipole, electric quadrupole and magnetic octupole hyperfine interaction matrix elements among $2s2p$ $^3\!P_{J}$ states. Based on these matrix elements, we evaluate the first-order, second-order and third-order hyperfine interaction parameters, and reanalyze the previously measurement of the hyperfine interval for the $2s2p$ $^3\!P_{J}$ state in $^9$Be by Blachman and Lurio. Detailed results and discussions are provided in the section~\ref{sec3}. The section~\ref{sec2} introduces the theory of hyperfine interaction and theory method. Finally, a summary is presented in the section~\ref{sec3}.

\section{THEORY}\label{sec2}
\subsection{Hyperfine Interaction}
The hyperfine interaction Hamiltonian can be expressed as
\begin{eqnarray}\label{eq1}
	H_{\rm {HFI}}=\sum _{k}{\textbf{\rm{T}}_{e}^{(k)}}\cdot \textbf{\rm{M}}_{n}^{(k)},
\end{eqnarray}
where $\textbf{\rm{T}}_{e}$ and $\textbf{\rm{M}}_{n}$ represent the spherical tensor operators of rank $k$ $(k>0)$ in the electronic and nuclear coordinate spaces, respectively. According to parity and the angular selection rules, $k$ must be even for electric moments and odd for magnetic moments. When the presence of the nucleus's multipolar fields is considered, the total electronic angular momentum $J$ is no longer conserved. The atomic angular momentum $\textbf{J}$ and the nuclear spin angular momentum $\textbf{I}$ couple to the total angular momentum $\textbf{F} = \textbf{I} + \textbf{J}$. The hyperfine state $|\gamma IJFM_F\rangle$ is constructed from
coupling a nuclear eigenstate $|\gamma IM_I\rangle$ with an atomic eigenstate $|\gamma IJM_J\rangle$ with $\gamma$ representing the remaining electronic quantum
numbers.

Compared with fine structure splitting, hyperfine splitting is much smaller, so the hyperfine interaction may be treated as a perturbation.
The hyperfine energy with the total angular momentum $F$ can be expressed as
\begin{small}
\begin{eqnarray}\label{eq2}
    W_{F} =W_{J}+\Delta{W_{F}^{(1)}}+\Delta{W_{F}^{(2)}}+\Delta{W_{F}^{(3)}}+\dots ,
\end{eqnarray}
\end{small}
where $W_{J}$ is the fine energy, and  $\Delta{W_{F}^{(n)}}$ represent the $n$-order corrections of hyperfine interaction to energy.
The first-order corrections of hyperfine interaction can be written as \cite{beloy2008hyperfine}:
\begin{small}
\begin{eqnarray}\label{eq3}
\Delta{W_F^{(1)}}&=&\langle\gamma IJFM_F|H_{\mathrm{HFI}}|\gamma IJFM_F\rangle
\nonumber \\
&=& \sum_{k}\limits (-1)^{I+J+F}\frac{\begin{Bmatrix} F & J & I\\k & I & J\end{Bmatrix}}{\begin{pmatrix}I & k & I\\-I & 0 & I\end{pmatrix}}\langle \gamma J||T_{e}^{(k)}||\gamma J\rangle \langle I\|M_n^{(k)}\|I\rangle.
\end{eqnarray}
\end{small}
The upper limit of $k$ is determined both by the electronic and the nuclear wave functions, and the matrix elements in Eq(\ref{eq3}) with respect to the quantum numbers $I$ and $J$ must vanish when $k > 2I$ or $k > 2J$. $\langle I\|M_n^{(k)}\|I\rangle$ and $\langle \gamma J||T_{e}^{(k)}||\gamma J\rangle$ are the nuclear and electric matrix element, respectively.
The nuclear matrix elements are given in terms of conventional nuclear moments: $\mu = \langle I\|M_n^{(1)}\|I\rangle$, $\frac{1}{2}Q = \langle I\|M_n^{(2)}\|I\rangle$, and $-\Omega= \langle I\|M_n^{(3)}\|I\rangle$. $\mu$, $Q$, and $\Omega$ are nuclear magnetic dipole, electric quadrupole, and magnetic octupole moments, respectively.  Restricted to $k\leq3$, the first-order corrections can be parameterized as:
\begin{small}
\begin{align}\label{eq4}
\Delta{W_F^{(1)}}=&\frac{1}{2}KA+\frac{1}{2}\frac{3K(K+1)-4I(I+1)J(J+1)}{2I(2I-1)2J(2J-1)}B
\nonumber \\
&+\frac{1}{[I(I-1)(2I-1)J(J-1)(2J-1)]} \times\left\{(5/4)K^3 \right.
\nonumber \\
& \left.+5K^2+K\times[-3I(I+1)\times J(J+1)+I(I+1) \right.
\nonumber \\
&\left.+J(J+1)+3]-5I(I+1)J(J+1)\right\}C,
\end{align}
\end{small}
where $K=F(F+1)-I(I+1)-J(J+1)$,  $A$ is the magnetic dipole HFS constant, $B$ represents the electric quadrupole HFS constant, and $C$ denotes the magnetic octupole HFS constant, which are defined as\cite{li2021ab}:
\begin{small}
\begin{eqnarray}
A=\frac{\mu}{I}\frac{\langle\gamma J\|T^{(1)}\|\gamma J\rangle}{\sqrt{J(J+1)(2J+1)}}
\end{eqnarray}
\end{small}
\begin{small}
\begin{eqnarray}
B=2Q\bigg[\frac{2J(2J-1)}{(2J+1)(2J+2)(2J+3)}\bigg]^{1/2}\langle\gamma J\|T^{(2)}\|\gamma J\rangle
\end{eqnarray}
\end{small}
\begin{small}
\begin{eqnarray}
C=\Omega\bigg[\frac{J(2J-1)(J-1)}{(J+1)(J+2)(2J+1)(2J+3)}\bigg]^{1/2}\langle\gamma J\|T^{(3)}\|\gamma J\rangle
\end{eqnarray}
\end{small}
The generalized second-order correction is \cite{beloy2008hyperfine,Li2023pra,Li2024pra}:
\begin{small}
\begin{align}\label{eq6}
W_F^{(2)}&=\sum\limits_{\gamma^{\prime}J^{\prime}}\frac{|\langle\gamma^{\prime}IJ^{\prime}FM_F|H_{\mathrm{HFI}}|\gamma IJFM_F\rangle|^2}{W_{\gamma J}-W_{\gamma^{\prime}J^{\prime}}}
\nonumber \\
&= \sum\limits_{\gamma^{'} J^{'} }^{}{\frac{1}{W_{\gamma J}-W_{\gamma^{'} J^{'}}  } }\sum_{k_1,k_2}^{}\begin{Bmatrix}F&J&I\\k_1&I&J^{'}\end{Bmatrix}\begin{Bmatrix}F&J&I\\k_2&I&J^{'}\end{Bmatrix}
\nonumber \\
&\quad\times \langle\gamma^{\prime}J^{\prime}\|T_e^{(k_1)}\|\gamma J\rangle\langle\gamma^{\prime}J^{\prime}\|T_e^{(k_2)}\|\gamma J\rangle
\nonumber \\
&\quad\times \langle I\|M_n^{(k_1)}\|I\rangle\langle I\|M_n^{(k_2)}\|I\rangle.
\end{align}
\end{small}
While the summation in Eq.(~\ref{eq6}) involves all possible excited electronic states, the vast majority of contributions come from magnetic dipole (M1) and electric quadrupole (E2) hyperfine interaction between two nearby fine-structure levels, i.e., $J^{'} = J\pm 1$, owing to small energy denominators. The second-order correction formulas of $2s2p$ $^3\!P_{1}$ and $2s2p$ $^3\!P_{2}$ are shown in Appendix A.

The generalized form of the third-order correction can be expressed as follows:
\begin{widetext}
\begin{small}
\begin{align}\label{eq7}
W_F^{(3)}&= \sum \limits_{\gamma^{\prime}J^{\prime}}\frac{|\langle\gamma^{\prime}IJ^{\prime}FM_F|H_{\mathrm{HFI}}|\gamma IJFM_F\rangle|^2}{\left ( W_{\gamma J}-W_{\gamma^{\prime}J^{\prime}}\right )^{2}}\times \left ( {\langle\gamma^{\prime}IJ^{\prime}FM_F|H_{\mathrm{HFI}}|\gamma^{\prime}IJ^{\prime}FM_F\rangle} -{\langle\gamma IJFM_F|H_{\mathrm{HFI}}|\gamma IJFM_F\rangle}\right )
\nonumber \\
& +  \sum \limits_{\gamma^{\prime\prime}J^{\prime\prime}}\frac{|\langle\gamma^{\prime\prime}IJ^{\prime\prime}FM_F|H_{\mathrm{HFI}}|\gamma IJFM_F\rangle|^2}{\left (W_{\gamma J}-W_{\gamma^{\prime\prime}J^{\prime\prime}}\right )^{2}}\times \left ( {\langle\gamma^{\prime\prime}IJ^{\prime\prime}FM_F|H_{\mathrm{HFI}}|\gamma^{\prime\prime}IJ^{\prime\prime}FM_F\rangle} -{\langle\gamma IJFM_F|H_{\mathrm{HFI}}|\gamma IJFM_F\rangle}\right )
\nonumber \\
& + \frac{2\langle\gamma^{\prime\prime}IJ^{\prime\prime}FM_F|H_{\mathrm{HFI}}|\gamma IJFM_F\rangle \langle\gamma^{\prime}IJ^{\prime}FM_F|H_{\mathrm{HFI}}|\gamma IJFM_F \rangle\langle\gamma^{\prime\prime}IJ^{\prime\prime}FM_F|H_{\mathrm{HFI}}|\gamma^{\prime} IJ^{\prime}FM_F\rangle}{\left ( W_{\gamma J}-W_{\gamma^{\prime}J^{\prime}}\right )\left ( E_{\gamma J}-E_{\gamma^{\prime\prime}J^{\prime\prime}}\right )},
\end{align}
\end{small}
\end{widetext}
As with the second-order correction, the primary contribution of the third-order correction originates from the M1 and E2 hyperfine interactions between adjacent fine-structure levels. The third-order correction formulas for the $2s2p$ $^3\!P_{1}$ and $2s2p$ $^3\!P_{2}$ states are provided in Appendix A.

For the $2s2p$ $^3\!P_1$  state,  the HFS constants $A$ and $B$ can be derived using the following formulas:
\begin{small}
\begin{align}\label{eq10}
A=&-\frac{1}{6} \delta W_{\frac{1}{2}-\frac{3}{2}}\left(^3\!P_1\right)-\frac{3}{10} \delta W_{\frac{3}{2}-\frac{5}{2}}\left(^3\!P_1\right)
\nonumber \\
&-\frac{1}{300}\chi +\frac{1}{180}\eta_{1}+\frac{1}{90}\eta +\frac{\sqrt{3}}{150}\zeta
\nonumber \\
&+\frac{1}{6}W_{\frac{1}{2}}^{\left(3\right)}\left(^3\!P_1\right)+\frac{2}{15}W_{\frac{3}{2}}^{\left(3\right)}\left(^3\!P_1\right)- \frac{3}{10}W_{\frac{5}{2}}^{\left(3\right)}\left(^3\!P_1\right),
\end{align}
\end{small}
\begin{small}
\begin{align}\label{eq11}
B=&\frac{1}{3} \delta W_{\frac{1}{2}-\frac{3}{2}}\left(^3\!P_1\right)-\frac{1}{5} \delta W_{\frac{3}{2}-\frac{5}{2}}\left(^3\!P_1\right)
\nonumber \\
&-\frac{1}{225}\eta+\frac{1}{45}\eta_{1}+\frac{\sqrt{3}}{150}\zeta
\nonumber \\
&-\frac{1}{3}W_{\frac{1}{2}}^{\left(3\right)}\left(^3\!P_1\right)+\frac{8}{15}W_{\frac{3}{2}}^{\left(3\right)}\left(^3\!P_1\right) -\frac{1}{5}W_{\frac{5}{2}}^{\left(3\right)}\left(^3\!P_1\right).
\end{align}
\end{small}
The HFS constants $A$ , $B$ and $C$ of the $2s2p$ $^3\!P_{2}$ state can be obtained using the following equations:
\begin{small}
\begin{align}\label{eq12}
A=&-\frac{3}{50} \delta W_{\frac{1}{2}-\frac{3}{2}}\left(^3\!P_2\right)-\frac{7}{50} \delta W_{\frac{3}{2}-\frac{5}{2}}\left(^3\!P_2\right)
\nonumber \\
&-\frac{4}{25} \delta W_{\frac{5}{2}-\frac{7}{2}}\left(^3\!P_2\right)+\frac{1}{300}\eta -\frac{1}{500\sqrt{3}}\zeta +\frac{1}{300}\chi
    \nonumber \\
&+\frac{3}{50}W_{\frac{1}{2}}^{\left(3\right)}\left(^3\!P_2\right)+\frac{4}{50}W_{\frac{3}{2}}^{\left(3\right)}\left(^3\!P_2\right)+\frac{1}{50}W_{\frac{5}{2}}^{\left(3\right)}\left(^3\!P_2\right),
\end{align}
\end{small}
\begin{small}
\begin{align}\label{eq13}
B=&\frac{2}{5} \delta W_{\frac{1}{2}-\frac{3}{2}}\left(^3\!P_2\right)+\frac{2}{5} \delta W_{\frac{3}{2}-\frac{5}{2}}\left(^3\!P_2\right)-\frac{16}{35} \delta W_{\frac{5}{2}-\frac{7}{2}}\left(^3\!P_2\right)
    \nonumber \\
&+\frac{2}{75}\eta +\frac{\sqrt{3}}{75}\zeta -\frac{2}{5}W_{\frac{1}{2}}^{\left(3\right)}\left(^3\!P_2\right) +\frac{6}{7}W_{\frac{5}{2}}^{\left(3\right)}\left(^3\!P_2\right),
\end{align}
\end{small}
\begin{small}
\begin{align}\label{eq14}
C=&-\frac{1}{50} \delta W_{\frac{1}{2}-\frac{3}{2}}\left(^3\!P_2\right)+\frac{1}{50} \delta W_{\frac{3}{2}-\frac{5}{2}}\left(^3\!P_2\right)
    \nonumber \\
&-\frac{1}{175} \delta W_{\frac{5}{2}-\frac{7}{2}}\left(^3\!P_2\right)+\frac{1}{500\sqrt{3}}\zeta
    \nonumber \\
&+\frac{1}{50}W_{\frac{1}{2}}^{\left(3\right)}\left(^3\!P_2\right)-\frac{1}{25}W_{\frac{3}{2}}^{\left(3\right)}\left(^3\!P_2\right) +\frac{9}{350}W_{\frac{5}{2}}^{\left(3\right)}\left(^3\!P_2\right).
\end{align}
\end{small}
Above, $\delta W_{F'-F}$ denotes the hyperfine splitting, which is given by $\delta W_{F'-F} = W_{F'} - W_{F}$. The hyperfine splitting values for the $2s2p$ $^3\!P_{1}$ and $2s2p$ $^3\!P_{2}$ states of $^9$Be are listed in Table~\ref{tab1}. The terms $\eta$, $\eta_{1}$, $\zeta$, and $\chi$ represent second-order corrections, while $W_{F}^{\left(3\right)}$ signifies the third-order correction. These parameters are defined in Appendix A and are provided through theoretical calculations.
\begin{table}[]
\caption[]{The hyperfine splitting values of the $2s2p$ $^3\!P_2$ and $2s2p$ $^3\!P_1$ states\cite{blachman1967hyperfine}.}
\begin{ruledtabular}
\begin{tabular}{lcc}
Hyperfine splitting & Values$(\rm MHz)$ \\
 \hline
$\delta W_{\frac{1}{2}-\frac{3}{2}}\left(^3\!P_2\right)$ &187.6157 $\pm0.0042$ \\
$\delta W_{\frac{3}{2}-\frac{5}{2}}\left(^3\!P_2\right)$ &312.0226 $\pm0.0021$ \\
$\delta W_{\frac{5}{2}-\frac{7}{2}}\left(^3\!P_2\right)$ &435.4773 $\pm0.0021$ \\
$\delta W_{\frac{1}{2}-\frac{3}{2}}\left(^3\!P_1\right)$ &202.9529 $\pm0.0015$ \\
$\delta W_{\frac{3}{2}-\frac{5}{2}}\left(^3\!P_1\right)$ &354.4365 $\pm0.0027$ \\
\end{tabular}\label{tab1}
\end{ruledtabular}
\end{table}
\subsection{Relativistic multiconfiguration methods}
In the present work, we utilize the MCDF and RCI methods within the GRASP2K framework to derive the wave-functions of the $2s2p$ $^3\!P_{J}$ in Be atom. These methods have been extensively outlined by Grant \cite{grant2007relativistic}. In the framework of multi-configuration methods, the atomic state function is expressed as an expansion of configuration state functions (CSFs):
\begin{small}
\begin{eqnarray}\label{eq20}
    \psi\left (\gamma J\pi \right )=\sum_{i}^{N_{\rm CSFS}} c_{i}\phi\left (\gamma _{i}J\pi \right ),
\end{eqnarray}
\end{small}
where $N_{\rm CSFS}$ denotes the total number of CSFs within the expansion, $J$ represents the total angular momentum,$\pi$ signifies the parity, and $\gamma_i$ is the additional quantum number to define each configuration state uniquely. The CSFs are characterized by $jj$-coupled antisymmetric products of Dirac orbitals. The expansion coefficient $c_i$ and the radial parts of the Dirac orbitals are determined through a self-consistent calculation process. Once a set of radial orbitals is established, configuration interaction can be executed, wherein the primary focus is on determining the expansion coefficients through diagonalization of the Hamiltonian matrix. In the RCI calculation, higher-order electron-electron interactions and the Breit interaction within the low-frequency limit approximation are also included. Additionally, minor corrections such as vacuum polarization (VP) and self-energy (SE) are incorporated into this computational framework.

For the calculation of the reduced matrix element, the expression is mainly in the following\cite{jonsson1996hfs92}:
\begin{small}
\begin{eqnarray}\label{eq21}
    \langle \gamma _{r}PJ||T_{e}^{(k)}||\gamma_{s}PJ' \rangle=\sum _{a,b}d_{ab}^{k}(rs)\langle n_{a}\kappa_{a}|| T_{e}^{(k)}|| n_{b}\kappa_{b}\rangle,
\end{eqnarray}
\end{small}
For the reduced matrix element, the expression of the operator $T_e^{(k)}$(k =1, 2) is as follows:
\begin{small}
\begin{eqnarray}\label{eq22}
\langle n_{a}\kappa_{a}|| T_{e}^{(1)}|| n_{b}\kappa_{b}\rangle  =&- \left (k_{a}+k_{b}\right )
\nonumber \\
& \times \langle -\kappa_{a}|| C^{1}|| \kappa_{b}\rangle\left [r^{-2}\right ]_{n_{a},\kappa_{a},n_{b}\kappa_{b}} \\\label{eq23}
\langle n_{a}\kappa_{a}|| T_{e}^{(2)}|| n_{b}\kappa_{b}\rangle  =&-\langle\kappa_{a}|| C^{2}|| \kappa_{b}\rangle\langle r^{-3}\rangle _{n_{a},\kappa_{a},n_{b}\kappa_{b}}
\end{eqnarray}
\end{small}
with
\begin{small}
\begin{eqnarray*}
\underline{}&\langle \kappa|| C^{k}|| \kappa'\rangle=\left (-1 \right )^{j_{\kappa}+\frac{1}{2} }\sqrt{(2j_{\kappa}+1)(2j_{\kappa'}+1)}\begin{pmatrix} j_{\kappa}&k &j_{\kappa'}\\\frac{1}{2} &0&-\frac{1}{2}\end{pmatrix} \pi \left (\ell_{\kappa},k,\ell_{\kappa'}  \right ) \\
&\pi \left (\ell_{\kappa},k,\ell_{\kappa'} \right ) =\begin{cases}1 \quad if\quad \ell_{\kappa}+k+\ell_{\kappa'} '\quad  \rm even, \\ 0 \quad \rm otherwise,\end{cases} \\
&[r^k]_{n\kappa n'\kappa'}=\int\limits_0^\infty r^k\big[P_{n\kappa}(r)Q_{n'\kappa'}(r)+Q_{n\kappa}(r)P_{n'\kappa'}(r)\big]dr,\\
&\langle r^k\rangle_{n\kappa n'\kappa'}=\int\limits_0^\infty r^k\big[P_{n\kappa}(r)P_{n'\kappa'}(r)+Q_{n\kappa}(r)Q_{n'\kappa'}(r)\big]dr.
\end{eqnarray*}
\end{small}

\section{Results and Discussion}\label{sec3}
\subsection{Active orbital sets}
For the Be atom, the $1s$ orbital constitutes the atomic core, while all other orbitals are classified as excited orbitals.  Table~\ref{tab2} outlines the active orbital sets employed in the present calculations. The active sets are denoted by specifying the highest orbital in each symmetry. For example, $\{4s,4p,4d,4f\}$ represents the orbital set encompassing $1s,2s,3s,4s,2p,3p,4p,3d,4d,4f$.
The single and double excitations from the reference configuration are referred to as SD excitations. The comprehensive set of single, double, triple, and quadruple excitations is denoted as SDTQ excitations. In the MCDF calculation, the extension of the CSFs is limited to SD excitations from the reference configuration. In the subsequent RCI calculation, in addition to the SD excitations of the CSFs expanded configuration to the maximum active orbital set, the subset of SDTQ excitations is also considered.
The calculation processes SD excitations to the maximum active orbital and SDTQ excitations to the subset orbitals by incrementally expanding layer-by-layer to the excitation orbitals.

\begin{table}[]
\caption[]{The active orbital sets in MCDF and RCI calculations. SD excitation represents the single and double excitation from the reference configuration, while
SDTQ excitation corresponding to the single, double, triple, and quadruple excitations from the reference configuration. }
\begin{ruledtabular}
\begin{tabular}{lcc}
Model & SD excitation & SDTQ excitation\\
 \hline
 1 & $\{4s,4p,4d,4f$\}           &$\{4s,4p,4d,4f$\}          \\
 2 & $\{5s,5p,5d,5f,5g$\}        &$\{5s,5p,5d,5f,5g$\}       \\
 3 & $\{6s,6p,6d,6f,6g,6h$\}     &$\{6s,6p,6d,6f,6g,6h$\}    \\
 4 & $\{7s,7p,7d,7f,7g,7h,7i$\}  &$\{6s,6p,6d,6f,6g,6h$\}    \\
 5 & $\{8s,8p,8d,8f,7g,7h,7i$\}  &$\{6s,6p,6d,6f,6g,6h$\}    \\
 6 & $\{9s,9p,9d,8f,7g,7h,7i$\}  &$\{6s,6p,6d,6f,6g,6h$\}    \\
 7 & $\{10s,10p,9d,8f,7g,7h,7i$\}&$\{6s,6p,6d,6f,6g,6h$\}    \\
 8 & $\{11s,10p,9d,8f,7g,7h,7i$\}&$\{6s,6p,6d,6f,6g,6h$\}    \\
 9 & $\{12s,10p,9d,8f,7g,7h,7i$\}&$\{6s,6p,6d,6f,6g,6h$\}    \\
\end{tabular}\label{tab2}
\end{ruledtabular}
\end{table}

\subsection{Energies and matrix elements}
\begin{table*}[]
\caption[]{The convergence of the total energy ( in a.u.) for the first eight states of the Be atom.}
\begin{ruledtabular}
\begin{tabular}{lcccccccccccccccc}
 Model & $2s_{2}$ $^1S_0$ &	$2s2p$ $^3\!P_0$ &	$2s2p$ $^3\!P_1$ &	$2s2p$ $^3\!P_2$ &	$2s2p$ $^1P_1$ & $2s3s$ $^3S_1$ & $2s3s$ $^1S_0$ & $2p_{2}$ $^1D_2$ \\
 \hline
1    & --14.6478532  & --14.5583458 & --14.5583415 & --14.5583278  & --14.4627217 & --14.4210753 & --14.4058223 & --14.3997720  \\
2    & --14.6599588  & --14.5634615 & --14.5634557 & --14.5634387  & --14.4680680 & --14.4253800 & --14.4118524 & --14.4050735  \\
3    & --14.6661326  & --14.5665337 & --14.5665288 & --14.5665139  & --14.4718425 & --14.4289088 & --14.4167993 & --14.4070028  \\
4    & --14.6680151  & --14.5678137 & --14.5678095 & --14.5677958  & --14.4733861 & --14.4307266 & --14.4186174 & --14.4088931  \\
5    & --14.6684215  & --14.5683108 & --14.5683074 & --14.5682953  & --14.4741152 & --14.4311817 & --14.4191028 & --14.4092776  \\
6    & --14.6685957  & --14.5685299 & --14.5685265 & --14.5685145  & --14.4744142 & --14.4313828 & --14.4194037 & --14.4094654  \\
7    & --14.6686464  & --14.5685769 & --14.5685737 & --14.5685622  & --14.4744852 & --14.4314295 & --14.4194554 & --14.4095075  \\
8    & --14.6686543  & --14.5685810 & --14.5685779 & --14.5685664  & --14.4744895 & --14.4314379 & --14.4194665 & --14.4095135  \\
9    & --14.6686564  & --14.5685822 & --14.5685790 & --14.5685675  & --14.4744911 & --14.4314431 & --14.4194786 & --14.4095143  \\
\end{tabular}\label{tab3}
\end{ruledtabular}
\end{table*}
Table~\ref{tab3} presents the convergence of the total energy ( in $a.u.$) for the first eight states of the Be atom. From table~\ref{tab3},  it is evident that the energy achieve five significant digits at the decimal point when the SD excitation are $\{$9s,9p,9d,8f,7g,7h,7i$\}$ and the SDTQ excitation are $\{$6s,6p,6d,6f,6g,6h$\}$.
Table~\ref{tab4} list the convergence of the excited energy(in $\rm cm^{-1}$) and compares these results with the experimental values from NIST database~\cite{NIST_ASD}. The last column ''Diff "($\%$) represents the relative difference between our final results and the NIST values. Table~\ref{tab4} shows that the relative difference does not exceed 0.1$\%$ for all states except the $2s2p$ $^1\!P_1$ state, where the relative difference is 0.12$\%$. It is also evident from table \ref{tab4} that the fine structure interval of $2s2p$ $^3\!P_{J}$ is quite small, suggesting that second-order corrections from hyperfine mixing between different $2s2p$ $^3\!P_{J}$ levels might be significant.
\begin{table*}[]
\caption[]{The convergence of the excited energy for the excited state of the Be atom ( in cm$^{-1}$) . The experimental values are from NIST database~\cite{NIST_ASD}.}
\begin{ruledtabular}
\begin{tabular}{lcccccccccccccccc}
 Model  &	$2s2p$ $^3\!P_0$ &	$2s2p$ $^3\!P_1$ &	$2s2p$ $^3\!P_2$ &	$2s2p$ $^1P_1$ & $2s3s$ $^3S_1$ & $2s3s$ $^1S_0$ & $2p_{2}$ $^1D_2$  \\
 \hline
1     & 19644.59   & 19645.53   & 19648.55   & 40631.66   	& 49771.99   	& 53119.64  	& 54447.52   \\
2     & 21178.71   & 21180.00   & 21183.73   & 42115.18   	& 51484.10   	& 54453.08  	& 55940.87   \\
3     & 21859.44   & 21860.50   & 21863.79   & 42641.75   	& 52064.62   	& 54722.34  	& 56872.41   \\
4     & 21991.66   & 21992.59   & 21995.59   & 42716.12   	& 52078.81   	& 54736.45  	& 56870.69   \\
5     & 21971.75   & 21972.51   & 21975.16   & 42645.30   	& 52068.12   	& 54719.14  	& 56875.50   \\
6     & 21961.91   & 21962.66   & 21965.29   & 42617.92   	& 52062.22    	& 54691.32  	& 56872.54   \\
7     & 21962.73   & 21963.42   & 21965.94   & 42613.45   	& 52063.08   	& 54691.10  	& 56874.41   \\
8     & 21963.54   & 21964.23   & 21966.75   & 42614.25   	& 52062.99   	& 54690.39  	& 56874.83   \\
9     & 21963.74   & 21964.43   & 21966.95   & 42614.35   	& 52062.29   	& 54688.19  	& 56875.11   \\
NIST~\cite{NIST_ASD}  & 21978.31   & 21978.93   & 21981.26   & 42565.45   	& 52080.94   	& 54677.35  & 56882.55  \\
Diff  & 0.066$\%$ & 0.066$\%$ & 0.065$\%$ & -0.12$\%$   & 0.036$\%$ 	& -0.020$\%$	& 0.013$\%$ \\
\end{tabular}\label{tab4}
\end{ruledtabular}
\end{table*}
\begin{table*}[]
\caption[]{Convergence for the M1 and E2 reduced matrix elements of the hyperfine interaction among different $2s2p$ $^3\!P_{J}$ levels.}
\begin{ruledtabular}
\begin{tabular}{lcccccccccccccccc}
\multirow{2}{*}{Modle}& \multicolumn{4}{c}{M1 reduced matrix elements}& & \multicolumn{4}{c}{E2 reduced matrix elements} \\
\cline{2-5} \cline{7-10}
 & $^3\!P_1$ $\rightarrow$ $^3\!P_0$ &  $^3\!P_1$ $\rightarrow$ $^3\!P_1$  & $^3\!P_2$ $\rightarrow$ $^3\!P_1$ & $^3\!P_2$ $\rightarrow$ $^3\!P_2$ &  & $^3\!P_1$ $\rightarrow$ $^3\!P_1$ & $^3\!P_2$ $\rightarrow$ $^3\!P_0$  & $^3\!P_2$ $\rightarrow$ $^3\!P_1$ & $^3\!P_2$ $\rightarrow$ $^3\!P_2$\\
\hline
1& 356.32  & 427.88  & --440.53 & 858.89 & & --31.37977   & --36.22734  & 54.33120  & 47.89783 \\
2& 354.07  & 423.19  & --435.41 & 854.21 & & --33.72371   & --38.93295  & 58.38828  & 51.47330 \\
3& 348.59  & 433.18  & --436.05 & 861.22 & & --38.31166   & --44.23611  & 66.35156  & 58.51115 \\
4& 353.49  & 435.17  & --440.09 & 868.98 & & --36.69466   & --42.36735  & 63.54592  & 56.03234 \\
5& 353.73  & 434.50  & --439.81 & 868.81 & & --37.44103   & --43.23015  & 64.84158  & 57.17765 \\
6& 353.15  & 434.15  & --439.34 & 867.60 & & --37.35441   & --43.12980  & 64.69057  & 57.04357 \\
7& 353.20  & 434.46  & --439.49 & 868.09 & & --37.43894   & --43.22744  & 64.83708  & 57.17286 \\
8& 353.16  & 434.43  & --439.45 & 868.02 & & --37.43890   & --43.22739  & 64.83701  & 57.17280 \\
9& 353.24  & 434.50  & --439.54 & 868.17 & & --37.43869   & --43.22715  & 64.83665  & 57.17248 \\
\end{tabular}\label{tab5}
\end{ruledtabular}
\end{table*}

Table~\ref{tab5} presents the convergence for the M1 and E2 reduced matrix elements of the hyperfine interaction among different $2s2p$ $^3\!P_{J}$ levels. Similar to the convergence observed in energy properties, these reduced matrix elements begin to converge when the SD excitation orbitals are  $\{$9s,9p,9d,8f,7g,7h,7i$\}$, and the SDTQ orbital are $\{$6s,6p,6d,6f,6g,6h$\}$. For the M1 reduced matrix elements, convergence is achieved to three significant figures, and for the E2 reduced matrix elements, convergence is achieved to five significant figures.
\begin{table*}
\caption[]{The extracted hyperfine-structure constants (in MHz) of $2s2p$ $^3\!P_{1}$ and $2s2p$ $^3\!P_{2}$ states in $^9$Be under considering
to first-order, second-order and third-order corrections. The numbers marked with asterisk are the extracted HFS constants (in MHz) of $2s2p$ $^3\!P_{2}$ state without considering the M3 hyperfine interaction. The values in parentheses are the uncertainties.}
\begin{ruledtabular}
\begin{tabular}{lcccccccccccc}
&\multicolumn{2}{c}{$^3\!P_1$} &\multicolumn{3}{c}{$^3\!P_2$} \\
\cline{2-3} \cline{5-7}
&  $A$  & $B$  &&  $A$  & $B$  & $C$  \\
\hline
First-order  & -140.15644 & -3.2364  && -124.61647 & 0.77800 & -0.00030371  \\
Second-order & -139.35669 & -0.7155  && -124.53370 & 1.45420 & 0.00003182   \\
Third-order  & -139.35661 & -0.7142  && -124.53409 & 1.45418 & 0.00011746   \\
             &            &          && -124.53738* & 1.4448* &              \\
             \\
Expt.~\cite{blachman1967hyperfine}&-139.373(12)&-0.753(44)&&-124.5368(17)&1.429(8)\\
Recommend    & -139.3566(83) &-0.714(30)&&-124.5341(10)&1.4542(67)&0.00012(8)\\
\end{tabular}\label{tab8}
\end{ruledtabular}
\end{table*}

With above M1 and E2 reduced matrix elements, we can subsequently calculate the first-order, second-order, and third-order corrections arising from the hyperfine interaction.
In these calculations, parameters such as the nuclear spin $I$, magnetic dipole moment $\mu$, and electric quadrupole moment $Q$ are required. For $^9$Be, the nuclear spin $I$ is $\frac{5}{2}$, and the magnetic dipole moment $\mu$ is $-1.177492$. The currently recommended reference value of the electric quadrupole moment $Q$ is $0.05288(38)$~b.  While the most recent few-body method calculations give an electric quadrupole moment of $Q=0.05350(14)$~b. The difference between these two values is less than $2\%$.
Our calculations indicate that a $2\%$ variation in the electric quadrupole moment $Q$ has a negligible impact on the second-order and third-order corrections and can be safely ignored. The parameters for second-order and third-order corrections are list in table~\ref{tab6} in MHz. From table~\ref{tab6}, one can find that
the primary contributions are from the M1-M1 terms of the second-order correction, followed by the M1-E2 terms of the second-order correction. The E2-E2 terms of second-order correction and third-order correction are significantly smaller compared to the aforementioned terms.

\begin{table}[]
\caption[]{The parameters for second-order and third-order corrections in MHz.}
\begin{ruledtabular}
\begin{tabular}{lcccccccccccc}
Simple &$^3\!P_1$ &$^3\!P_2$\\
\hline
$\eta$   &\multicolumn{2}{c}{25.0315}\\
$\zeta$  & \multicolumn{2}{c}{0.29058}  \\
$\chi$   & \multicolumn{2}{c}{0.000843}\\
$\eta_1$ & 59.1461   &        \\
$W_{\frac{1}{2}}^{\left(3\right)}$& 0.001605 & -0.001605  \\
$W_{\frac{3}{2}}^{\left(3\right)}$& 0.003761 & -0.003440  \\
$W_{\frac{5}{2}}^{\left(3\right)}$& 0.000772 & -000772    \\
$W_{\frac{7}{2}}^{\left(3\right)}$& 0        &  0         \\
\end{tabular}\label{tab6}
\end{ruledtabular}
\end{table}

\subsection{Updated hyperfine-structure constants}
According to the Eqs.(\ref{eq10}-\ref{eq14}), the hyperfine-structure constants of $2s2p$ $^3\!P_{1}$ and $2s2p$ $^3\!P_{2}$ states in $^9$Be
can be extracted by combining the previous measurement of the hyperfine interval for the $2s2p$ $^3\!P$ state~\cite{blachman1967hyperfine} and
the calculated second-order and third-order corrections.

Table~\ref{tab8} lists the extracted hyperfine structure constants for $2s2p$ $^3\!P_{1}$ and $2s2p$ $^3\!P_{2}$ states in $^9$Be, considering first-order, second-order, and third-order corrections, respectively. From table\ref{tab8}, it can be observed that for both states, the second-order correction contributes significantly to the HFS $B$. For the $^3\!P_{1}$ state, the contribution of the second-order correction has an opposite sign to that of the first-order correction, leading to a significant cancellation between them. This indicates that the HFS $B$ for the $^3\!P_{1}$ state is highly sensitive to second-order correction. Our data analysis shows that HFS $B$ for the $^3\!P_{1}$ state is extremely sensitive to the accuracy for the $^3\!P_{1}$-$^3\!P_{0}$ transition matrix element; a 1\% change in this transition matrix element causes a 7\% change in HFS $B$. Therefore, to accurately extract the HFS $B$ for the $^3\!P_{1}$ state, high precision in the transition matrix elements is required. This also implies that extracting the nuclear electric quadrupole moment $Q$ from the HFS $B$ of the $^3\!P_{1}$ state may not be reliable. For the $^3\!P_{2}$ state, the second-order correction has the same sign as the first-order correction and is very close in magnitude. The third-order corrections for the  HFS $A$ and $B$ of the $^3\!P_{1}$ and $^3\!P_{2}$ states are small and can be almost ignored within the current precision. However, for HFS $C$ of the $^3\!P_{2}$ state, the second-order correction and the first-order correction have opposite signs, with the second-order value slightly larger than the first-order value, resulting in significant cancellation. The HFS $C$ after considering the second-order correction is nearly an order of magnitude smaller than the first-order result but has the opposite sign. The third-order correction has the same sign as the second-order correction, and its magnitude is about one-third that of the second-order correction. Due to the significant cancellation between the first-order and second-order corrections, the contribution of the third-order correction becomes more important. This result implies that to accurately extract the HFS $C$, both second-order and third-order corrections need to be considered.

Table~\ref{tab8} also list the previously reported values~\cite{blachman1967hyperfine}, which considered second-order and third-order correction values but ignored the M3 hyperfine interaction. To directly compare with the previously reported values, we provide the extracted HFS $A$ and $B$ for the $^3\!P_{2}$ state without considering the M3 hyperfine interaction. These two values are marked with asterisks. Comparing with the previously reported values, it can be observed that our HFS $A$ for both the $^3\!P_{1}$ and $^3\!P_{2}$ states are consistent with the experimental values, but there are discrepancies in the HFS $B$; our HFS $B$ for the $^3\!P_{1}$ state is about 5\% smaller than the previously reported values, although it is still within the uncertainty. Our HFS $B$ for the $^3\!P_{2}$  state is about 1.1\% larger than the previously reported values. The discrepancy in the HFS $B$ is due to the differences in the off-diagonal hyperfine interaction matrix elements in the second-order correction calculations. Additionally, by comparison, we observe that the M3 hyperfine interaction contributes about 0.65\% to the HFS $B$ of the $^3\!P_{2}$  state. After considering the second-order and third-order corrections, the HFS $B$ value of the  $^3\!P_{2}$ state is about 1.7\% larger than the previously reported values.

To assess the accuracy of our theoretical calculations, we first directly compare the theoretical and experimental values of the HFS $A$. Using the reduced matrix elements from table~\ref{tab4}, the HFS $A$ for the $^3\!P_{1}$ and $^3\!P_{2}$ states are calculated to be 139.2 MHz and 124.4 MHz, respectively, with differences from the updated HFS constant $A$ of 0.13\% and 0.12\%. This indicates that the accuracy of our M1 reduced matrix elements is high, better than 0.5\%. The accuracy of our E2 reduced matrix elements can be assessed
 by comparing with the results from a recent high-precision few-body precision calculation~\cite{puchalski2021hyperfine}. Our calculated $B/Q$ value for the $^3\!P_{2}$ state is $27.333$ MHz, which is 0.69\% larger than the value of $27.14887(3)$ MHz from few-body precision calculation~\cite{puchalski2021hyperfine}. Therefore, we conservatively estimate that the accuracy of our M1 and E2 reduced matrix elements is 0.5\%. Considering both the experimental measurement precision and our theoretical calculation accuracy, we provide the final recommended values and their corresponding uncertainties, which are listed in the last row of the table~\ref{tab8}. The uncertainties for HFS $A$  and $B$ primarily arise from theoretical calculations, while the uncertainty for HFS $C$ mainly stems from experimental measurements.

\subsection{Nuclear electric quadrupole moment $Q$}

Combining the theoretical calculation values of $B/Q$ and undated HFS $B$ of the $^3\!P_{2}$ state obtained in this work, we can extract the electric quadrupole moment $Q$ of the $^9$Be nucleus. Based on our theoretical calculation result of $B/Q=27.333$ MHz, the extracted Q value is equal to $0.05320(50)$~b, considering the uncertainty caused by both theoretical calculations and experimental measurements. As mentioned in the introduction, the electric quadrupole moment $Q$ determined by previous many-body calculations~\cite{ray1973study,sinano1973hfs,beck1984fine,sundholm1991large,jonsson1993large,nemouchi2003theoretical,beloy2008hyperfine} were all based on previously reported result of HFS $B=1.429(8)$ MHz of the $^3\!P_{2}$~\cite{blachman1967hyperfine}. We applied the undated HFS $B$ of the $^3\!P_{2}$ state, combined with early many-body results, to redetermine $Q$. The early and newly determined values $Q$ are listed in table~\ref{tab9}. It can be found that all newly determined $Q$  are around $0.0535$~b, with a difference of less than 1\%. All results are consistent with the high-precision values obtained by the few-body precision calculation~\cite{puchalski2021hyperfine} considering the uncertainty, with a difference of less than 1\%. This indicates that the inaccuracy of the previously reported HFS $B$ of the $^3\!P_{2}$ state is the main reason for the difference between the $Q$ values determined by the many-body calculations and the one obtained by the few-body precision calculation.  On other words, many-body methods such as multiconfiguration Hartree-Fock, multiconfiguration Dirac-Fock, configuration interaction, and coupled cluster methods, are capable of accurately calculating hyperfine interaction properties of Be.
\begin{small}
\begin{table}[]
\caption[]{Comparison of the electric quadrupole moment $Q$ (in b) of the $^9$Be nucleus. The early and newly determined values are marked by $Q(\rm old)$ and $Q(\rm new)$, respectively. The values in parentheses are the uncertainties. }
\begin{ruledtabular}
\begin{tabular}{llllllllllllllll}
Source                                                   &$Q(\rm old)$    &$Q(\rm new)$\\
\hline
Experiment\cite{blachman1967hyperfine}                   & 0.049(3)        &             \\
Ray \textit{ et. al.}\cite{ray1973study}                 & 0.0525(3)       & 0.0534(3)   \\
Sundholm and Olsen\cite{sundholm1991large}               & 0.05288(38)     & 0.05381(39) \\
J{\"o}nsson and Fisher\cite{jonsson1993large}            & 0.05256         & 0.05349(25)     \\
Nemouchi \textit{ et. al.}\cite{nemouchi2003theoretical} & 0.05277         & 0.05370(25)     \\
Beloy \textit{ et. al.}\cite{beloy2008hyperfine}         & 0.053(3)        &             \\
\\
Puchalski \textit{ et. al.}\cite{puchalski2021hyperfine} & \multicolumn{2}{c}{0.05350(14)} \\
This work                                                & \multicolumn{2}{c}{0.05320(50)}   \\
\end{tabular}\label{tab9}
\end{ruledtabular}
\end{table}
\end{small}

\section{Conclusion}\label{sec4}

In this work, we employ MCDF method to calculate hyperfine-structure properties of  the $2s2p$ $^3\!P_{J}$ state in $^9$Be. The hyperfine-structure properties include first-order hyperfine interaction parameters, along with second-order and third-order corrections due to the hyperfine mixing of different $2s2p$ $^3\!P_{J}$ levels. Leveraging these theoretical results, we reanalyze the previously reported measurement of the hyperfine interval for the $2s2p$ $^3\!P_{J}$ state in $^9$Be by Blachman and Lurio. Our analysis indicates that the second-order correction notably influences the HFS $B$ of both the $^3\!P_{1}$ and $^3\!P_{2}$ states. Specifically, the HFS $B$ for the $^3\!P_{1}$ state is highly sensitive to second-order correction. The third-order corrections to the  HFS $A$ and $B$ of the $^3\!P_{1}$ and $^3\!P_{2}$ states are minimal and can be largely disregarded within the current precision. Conversely, both the second-order and third-order corrections contribute significantly to the HFS $C$ of the $^3\!P_{2}$ state. Considering both the experimental measurement precision and our theoretical calculation accuracy, we provide the final recommended values and the corresponding uncertainties for the HFS constants for the $2s2p$ $^3\!P_{1}$  and $2s2p$ $^3\!P_{2}$ states in $^9$Be. Notably, the undated HFS $B$ for the $2s2p$ $^3\!P_{2}$ state is approximately 1.7\% larger then the previously reported value. Subsequently, we extract the electric quadrupole moment $Q$ of $^9$Be nucleus to be 0.05320(50) b  by integrating our theoretical results with the updated HFS $B$ for the $2s2p$ $^3\!P_{2}$ state. Additionally, we redetermine the electric quadrupole moment $Q$ by combining early many-body calculation results and the undated HFS $B$ obtained in this work. All newly determined $Q$ are around $0.0535$~b.  These values are consistent with high-precision values derived from few-body precision calculation~\cite{puchalski2021hyperfine}, considering the uncertainty. The discrepancy between $Q$ values obtained through early many-body calculations and recent few-body precision calculation can be attributed to the inaccuracy of the previously reported HFS $B$ constant for the $2s2p$ $^3\!P_{2}$ state.

\begin{acknowledgments}
We are grateful to Prof. T.-Y. Shi and Dr. C.-X. Song for reading our manuscript. The work was supported by the National Natural Science Foundation of China under Grant No.12174268.
\end{acknowledgments}
\begin{widetext}
\section*{Appendix A1: the second-order and third-order correction formulas for $~^3\!P_1$and $~^3\!P_2$}
\setcounter{equation}{0} \renewcommand{\theequation}{A\arabic{equation}}
The formulas for second-order corrections of $^3\!P_{1}$ and $^3\!P_{2}$ are as follows:

\begin{align}\label{eqA1}
W_{F}^{(2)}\left(~^3\!P_2\right)=&\left | \begin{Bmatrix} F & 2 & I\\ 1 & I & 1\end{Bmatrix} \right | ^{2}\eta+\left | \begin{Bmatrix} F & 2 & I\\ 2 & I & 1\end{Bmatrix} \right | ^{2}+\begin{Bmatrix} F & 2 & I\\ 1 & I & 1\end{Bmatrix}\times\begin{Bmatrix} F & 2 & I\\ 2 & I & 1\end{Bmatrix}\zeta,  \\\label{eqA2}
W_{F}^{(2)}\left(~^3\!P_1\right) =& \left | \begin{Bmatrix} F & 1 & I\\ 1 & I & 0\end{Bmatrix} \right | ^{2}\eta_1 - \left | \begin{Bmatrix} F & 1 & I\\ 1 & I & 2\end{Bmatrix} \right | ^{2}\eta -\left | \begin{Bmatrix} F & 1 & I\\ 2 & I & 2\end{Bmatrix} \right | ^{2}\chi  -\begin{Bmatrix} F & 1 & I\\ 1 & I & 2\end{Bmatrix} \times\begin{Bmatrix} F & 1 & I\\ 2 & I & 2\end{Bmatrix}\zeta ,
\end{align}
with
\begin{small}
\begin{equation*}
\begin{aligned}
\eta&=\frac{(I+1)(2I+1)}I\mu^2\frac{|\langle~^3\!P_2||T_e^{(1)}||~^3\!P_1\rangle |^2}{E_{~^3\!P_2}-E_{~^3\!P_1}}, \\
\zeta&=\frac{(I+1)(2I+1)}{I}\sqrt{\frac{2I+3}{2I-1}} \quad\times\mu Q\frac{\langle~^3\!P_2||T_e^{(1)}||~^3\!P_1\rangle\langle~^3\!P_2||T_e^{(2)}||~^3\!P_1\rangle}{E_{~^3\!P_2}-E_{~^3\!P_1 }},\\
\chi &=\frac{(I+1)(2I+1)(2I+3)}{I(2I-1)}\times\frac{Q^{2}}{4} \frac{\left |  \langle~^3\!P_2||T_e^{(2)}||~^3\!P_1\rangle  \right |^{2}}{E_{~^3\!P_2}-E_{~^3\!P_1}}, \\
\eta_{1}&=\frac{(I+1)(2I+1)}I\mu^2\frac{|\langle~^3\!P_1||T_e^{(1)}||~^3\!P_0\rangle|^2}{E_{~^3\!P_1}-E_{~^3\!P_0}}, \\
\end{aligned}
\end{equation*}
\end{small}

The formulas for third-order corrections of $^3\!P_{1}$ and $^3\!P_{2}$ were derived respectively. Their expressions are as follows:

\begin{small}
\begin{align}\label{eqA3}
W_{F}^{(3)}\left(^3\!P_2\right)  =& \left ( \left | \begin{Bmatrix}F & J &I \\1 & I &J^{\prime }\end{Bmatrix} \right |^{2}\eta^{\prime} + \left | \begin{Bmatrix}F & J &I \\2 & I &J^{\prime }\end{Bmatrix}  \right |^{2}\chi_{1}^{\prime} +  \begin{Bmatrix}F & J &I \\1 & I &J^{\prime }\end{Bmatrix} \begin{Bmatrix}F & J &I \\2 & I &J^{\prime }\end{Bmatrix}\zeta^{\prime}    \right )
\nonumber \\
& \times \left [ \left ( \begin{Bmatrix}F & J^{\prime } &I \\1 & I &J^{\prime }\end{Bmatrix}X_{1}+\begin{Bmatrix}F & J^{\prime } &I \\2 & I &J^{\prime }\end{Bmatrix}X_{2} \right )- \left ( \begin{Bmatrix}F & J &I \\1 & I &J\end{Bmatrix}X_{3}+\begin{Bmatrix}F & J &I \\2 & I &J\end{Bmatrix}X_{4} \right ) \right ]
\nonumber \\
& + \left | \begin{Bmatrix}F & J &I \\2 & I &J^{\prime\prime  }\end{Bmatrix} \right |^{2}\chi_{2}^{\prime } \times  \left [ -\left ( \begin{Bmatrix}F & J &I \\1 & I &J\end{Bmatrix}X_{3}+\begin{Bmatrix}F & J &I \\2 & I &J\end{Bmatrix}X_{4} \right )  \right ]
\nonumber \\
& + \begin{Bmatrix}F & J &I \\1 & I &J^{\prime }\end{Bmatrix} \begin{Bmatrix}F & J^{\prime } &I \\1 & I &J^{\prime\prime  }\end{Bmatrix} \begin{Bmatrix}F & J &I \\2 & I &J^{\prime\prime  }\end{Bmatrix}X_{5}+ \begin{Bmatrix}F & J &I \\2 & I &J^{\prime }\end{Bmatrix} \begin{Bmatrix}F & J^{\prime } &I \\1 & I &J^{\prime\prime  }\end{Bmatrix} \begin{Bmatrix}F & J &I \\2 & I &J^{\prime\prime  }\end{Bmatrix}X_{6} \\\label{eqA4}
W_{F}^{(3)} \left(^3\!P_1\right) =& \left ( \left | \begin{Bmatrix}F & J &I \\1 & I &J^{\prime }\end{Bmatrix} \right |^{2}\eta^{\prime} + \left | \begin{Bmatrix}F & J &I \\2 & I &J^{\prime }\end{Bmatrix}  \right |^{2}\chi_{1}^{\prime} +  \begin{Bmatrix}F & J &I \\1 & I &J^{\prime }\end{Bmatrix} \begin{Bmatrix}F & J &I \\2 & I &J^{\prime }\end{Bmatrix}\zeta^{\prime}    \right )
\nonumber \\
& \times \left [ \left ( \begin{Bmatrix}F & J^{\prime } &I \\1 & I &J^{\prime }\end{Bmatrix}X_{1}+\begin{Bmatrix}F & J^{\prime } &I \\2 & I &J^{\prime }\end{Bmatrix}X_{2} \right )- \left ( \begin{Bmatrix}F & J &I \\1 & I &J\end{Bmatrix}X_{3}+\begin{Bmatrix}F & J &I \\2 & I &J\end{Bmatrix}X_{4} \right ) \right ]
\nonumber \\
& + \left | \begin{Bmatrix}F & J &I \\1 & I &J^{\prime\prime  }\end{Bmatrix} \right |^{2}\eta_{1}^{\prime } \times  \left [ -\left ( \begin{Bmatrix}F & J &I \\1 & I &J\end{Bmatrix}X_{3}+\begin{Bmatrix}F & J &I \\2 & I &J\end{Bmatrix}X_{4} \right )  \right ]
\nonumber \\
& + \begin{Bmatrix}F & J &I \\1 & I &J^{\prime }\end{Bmatrix} \begin{Bmatrix}F & J^{\prime } &I \\1 & I &J^{\prime\prime  }\end{Bmatrix} \begin{Bmatrix}F & J &I \\2 & I &J^{\prime\prime  }\end{Bmatrix}X_{5}+ \begin{Bmatrix}F & J &I \\2 & I &J^{\prime }\end{Bmatrix} \begin{Bmatrix}F & J^{\prime } &I \\1 & I &J^{\prime\prime  }\end{Bmatrix} \begin{Bmatrix}F & J &I \\2 & I &J^{\prime\prime  }\end{Bmatrix}X_{6}
\end{align}
\end{small}
with
\begin{equation*}
\begin{aligned}
\eta^{\prime}&=\frac{(I+1)(2I+1)}I\mu^2\frac{|\langle\gamma J^{\prime}||T_e^{(1)}||\gamma J\rangle|^2}{\left(E_{\gamma J}-E_{\gamma J^{\prime}}\right)^{2}} \\
\eta_{1}^{\prime}&=\frac{(I+1)(2I+1)}I\mu^2\frac{|\langle\gamma J^{\prime\prime}||T_e^{(1)}||\gamma J\rangle|^2}{\left(E_{\gamma J}-E_{\gamma J^{\prime\prime}}\right)^{2}} \\
\zeta^{\prime}&=\frac{(I+1)(2I+1)}{I}\sqrt{\frac{2I+3}{2I-1}}\times\mu Q\frac{\langle\gamma J^{\prime}||T_e^{(1)}||\gamma J\rangle\langle\gamma J^{\prime}||T_e^{(2)}||\gamma J\rangle}{\left(E_{\gamma J}-E_{\gamma J^{\prime}}\right)^{2}} \\
\chi_{1}^{\prime} &=\frac{(I+1)(2I+1)(2I+3)}{I(2I-1)}\times\frac{Q^{2}}{4} \frac{\left |  \langle\gamma J^{\prime}||T_e^{(2)}||\gamma J\rangle  \right |^{2}}{\left(E_{\gamma J}-E_{\gamma J^{\prime}}\right)^{2}} \\
\chi_{2}^{\prime} &=\frac{(I+1)(2I+1)(2I+3)}{I(2I-1)}\times\frac{Q^{2}}{4} \frac{\left |  \langle\gamma J^{\prime\prime}||T_e^{(2)}||\gamma J\rangle  \right |^{2}}{\left(E_{\gamma J}-E_{\gamma J^{\prime\prime}}\right)^{2}} \\
X_{1}&=(-1)^{I+J+F}\frac {\mu \sqrt{I(I+1)(2I+1)} }{I}\langle\gamma J^{\prime}||T_e^{(1)}||\gamma J^{\prime}\rangle \\
X_{2}&=(-1)^{I+J+F}\frac {Q \sqrt{I(I+1)(2I+1)(2I+3)(2I-1)} }{2I(2I-1)}\langle\gamma J^{\prime}||T_e^{(2)}||\gamma J^{\prime}\rangle \\
X_{3}&=(-1)^{I+J+F}\frac {\mu \sqrt{I(I+1)(2I+1)} }{I}\langle\gamma J||T_e^{(1)}||\gamma J\rangle \\
X_{4}&=(-1)^{I+J+F}\frac {Q \sqrt{I(I+1)(2I+1)(2I+3)(2I-1)} }{2I(2I-1)}\langle\gamma J||T_e^{(2)}||\gamma J\rangle  \\
X_{5}&=(-1)^{I+J^{\prime}+F}\frac {\mu^{2}Q(I+1)(2I+1) \sqrt{I(I+1)(2I+1)(2I+3)(2I-1)} }{I^{2}\left(2I-1\right)}
 \times \frac{\langle\gamma J^{\prime}||T_e^{(1)}||\gamma J\rangle\langle\gamma J^{\prime\prime}||T_e^{(1)}||\gamma J^{\prime}\rangle\langle\gamma J^{\prime\prime}||T_e^{(2)}||\gamma J\rangle}{\left(E_{\gamma J}-E_{\gamma J^{\prime}}\right)\left(E_{\gamma J}-E_{\gamma J^{\prime\prime}}\right)}  \\
X_{6}&=(-1)^{I+J^{\prime}+F}\frac {\mu Q^{2}(I+1)(2I+1)(2I+3) \sqrt{I(I+1)(2I+1)} }{2I^{2}(2I-1)}
 \times \frac{\langle\gamma J^{\prime}||T_e^{(2)}||\gamma J\rangle\langle\gamma J^{\prime\prime}||T_e^{(1)}||\gamma J^{\prime}\rangle\langle\gamma J^{\prime\prime}||T_e^{(2)}||\gamma J\rangle}{\left(E_{\gamma J}-E_{\gamma J^{\prime}}\right)\left(E_{\gamma J}-E_{\gamma J^{\prime\prime}}\right)}
\end{aligned}
\end{equation*}
where $J^{\prime}=J+1$ and $J^{\prime\prime}=J-1$.
\end{widetext}

%

\end{document}